# Atomic-layer-resolved composition and electronic structure of the cuprate $Bi_2Sr_2CaCu_2O_{8+\delta}$ from soft x-ray standing-wave photoemission

Cheng-Tai Kuo,[1,2,*] Shih-Chieh Lin,[1,2] Giuseppina Conti,[1,2] Shu-Ting Pi,[1] Luca Moreschini,[3] Aaron Bostwick,[3] Julia Meyer-Ilse,[2] Eric Gullikson,[2] Jeffrey B. Kortright,[2] Slavomir Nemšák,[3,4] Julien E. Rault,[5] Patrick Le Fèvre,[5] François Bertran,[5] Andrés F. Santander-Syro,[6] Ivan A.Vartanyants,[7,8] Warren E. Pickett,[1] Romuald Saint-Martin,[9] Amina Taleb,[5] and Charles S. Fadley[1,2,†]

[1]*Department of Physics, University of California Davis, Davis, California 95616, USA*
[2]*Materials Sciences Division, Lawrence Berkeley National Laboratory, Berkeley, California 94720, USA*
[3]*Advanced Light Source, Lawrence Berkeley National Laboratory, Berkeley, California 94720, USA*
[4]*Peter Grünberg Institut PGI-6, Research Center Jülich, 52425 Jülich, Germany*
[5]*Synchrotron SOLEIL, L'Orme des Merisiers, Saint-Aubin-BP48, 91192 Gif-sur-Yvette, France*
[6]*CSNSM, Université Paris-Sud, CNRS/IN2P3, Université Paris-Saclay, 91405 Orsay Cedex, France*
[7]*Deutsches Elektronen-Synchrotron DESY, Notkestraße 85, D-22607 Hamburg, Germany*
[8]*National Research Nuclear University MEPhI (Moscow Engineering Physics Institute), Kashirskoe Shosse 31, Moscow, 115409, Russia*
[9]*SP2M-ICMMO-UMR-CNRS 8182, Université Paris-Sud, Université Paris-Saclay, 91405 Orsay Cedex, France*
[*] Corresponding author: chengtaikuo@lbl.gov
[†] Corresponding author: fadley@lbl.gov

## ABSTRACT

A major remaining challenge in the superconducting cuprates is the unambiguous differentiation of the composition and electronic structure of the $CuO_2$ layers and those of the intermediate layers. The large c axis for these materials permits employing soft x-ray (930.3 eV) standing wave (SW) excitation in photoemission that yields atomic layer-by-atomic layer depth resolution of these properties. Applying SW photoemission to $Bi_2Sr_2CaCu_2O_{8+\delta}$ yields the depth distribution of atomic composition and the layer-resolved densities of states. We detect significant Ca presence in the SrO layers and oxygen bonding to three different cations. The layer-resolved valence electronic structure is found to be strongly influenced by the atomic supermodulation structure--as determined by comparison to density functional theory calculations, by Ca-Sr intermixing, and by correlation effects associated with the Cu 3d-3d Coulomb interaction, further clarifying the complex interactions in this prototypical cuprate. Measurements of this type for other quasi-two-dimensional materials with large-c represent a promising future direction.

**I. INTRODUCTION**

The cuprate high-temperature superconductors have attracted much attention and been extensively studied, but are still not fully understood. It is believed that superconductivity is related to hole- or electron- doping within their layered quasi-2D crystallographic structures, with the key element being the $CuO_2$ planes [1,2]. Characterizing this basic element in superconductivity is thus critical, and some important challenges remaining are to differentiate the electronic structure of the $CuO_2$ layers from those of the intermediate layers, as well as the elemental composition of each layer.

Photoemission spectroscopy, especially angle-resolved photoemission spectroscopy (ARPES), is one of the most powerful techniques for visualizing the electronic structure in materials [1,3,4]. Conventional ARPES measurements are performed with excitation energies of ~20 to 150 eV that yield high surface sensitivity due to the short electron inelastic mean-free paths (IMFP, $\lambda_{IMFP}$) [5] of ~3-6 Å. For materials with small unit-cell dimensions perpendicular to the surface and inert, easily cleavable or *in situ*-preparable surfaces, ARPES can provide unique information on properties close to that of bulk. However, for materials with large c-axis parameters, e.g. the cuprates, it can be argued that conventional ARPES preferentially samples the topmost atomic layers rather than the full unit cell. For example, in the case of $Bi_2Sr_2CaCu_2O_{8+\delta}$ (Bi2212), the c-axis parameter is ~30.7 Å, and its first $CuO_2$ layer in the unit cell is ~6 Å below the cleaved surface; thus, for conventional ARPES, the contributions from the first $CuO_2$ layer will be attenuated by ~ $e^{-1}$ = 0.37, and they will be even more extreme for the deeper layers. Beyond this, in conventional ARPES, the only way to distinguish element-specific behavior is to use resonant photoemission that would selectively enhance the different layer contributions [6,7]. But quantitative interpretation of resonant photoemission is difficult, and the number of elements that can be studied is limited by the suitable core levels to excite resonantly. Standing-wave (SW) photoemission provides a method to get around these limitations of conventional ARPES and resonant photoemission.

X-ray SW excitation with energy ~2-10 keV in connection with spectroscopy was introduced some time ago [8], and its theory and applications have been reviewed in detail [9]. SW *hard x-ray* photoemission at a few keV has been used to derive the spatial distribution of composition and differentiate the element-specific matrix-element weighted densities of states (DOSs) within the unit cells of several solids [10,11,12], including $YBa_2Cu_3O_{7-\delta}$ (YBCO)[13]. For higher photon energies above the ca. 1 keV regime, these DOSs can be considered to be weighted by differential atomic-cross sections, and it is at this level that we will analyze our data.

In this work, we have chosen *soft x-ray* photoemission to study Bi2212, utilizing its (002) Bragg reflection to generate the SW. A photon energy of 930.3 eV was further chosen near an absorption resonance to maximize the SW strength; see the detail in Supplemental Material S2 [14]. The IMFP for Bi2212 at the excitation energy of ~930 eV calculated from the TPP-2M formula [5] is ~1.5 nm. Given that the intensity of collected photoelectrons decays as exp×-(L/ $\lambda_{IMFP}$), where L is the depth, ~99% of the collected photoelectrons are from the top 3 unit cells of Bi2212. For the excitation energies of 20-150 eV, ~99% of the collected photoelectrons are from only the first unit cell of Bi2212. Therefore, the excitation energy of ~930 eV is relatively sufficient to determine the bulk electronic structure. With the lower energy, it enables higher energy resolution and greater sensitivity to electron momentum than with a higher multi-keV energy. Choosing the soft x-ray of ~930 eV thus means simultaneously having higher reflectivity, sufficient bulk sensitivity, reasonable energy resolution, and better sensitivity to electron momentum, making it superior to hard x-ray excitation for our SW-XPS study.

## II. EXPERIMENTAL AND COMPUTATIONAL DETAILS

The Bi2212 single crystal in this work is optimally doped, and the critical temperature ($T_c$) determined by SQUID is ~93K. Details concerning the sample growth and characterization are in Supplemental Material S1 [14]. X-ray reflectivity measurements were performed at beamline 6.3.2 of Advanced Light Source (ALS). The Cr absorption edge (574.1 eV) is used for the energy calibration. ARPES, x-ray absorption spectroscopy (XAS) and SW photoemission (or SW-XPS) measurements were performed at beamline 7.0.2 (MAESTRO) of ALS, and the beamline CASSIOPEE of SOLEIL. The SW-XPS measurement was carried out at ~77K, at which the sample was superconducting. Note that the SW-XPS measurement could also be carried out in the normal state, as the temperature and the very small gap at 77K should have very small impact on the experimental results at this energy resolution.

The reflectivity and SW rocking curve data were analyzed using SW theory based on dynamical x-ray diffraction. The resonant Cu atomic scattering factors were calculated from a Cu $L_3$ XAS spectrum using Kramers-Kronig relations. The atomic coordinates of Bi2212 with supermodulation were obtained from Ref. [24]. The electronic structures were calculated using the first-principles package Quantum Espresso [25] with generalized gradient approximation (GGA) [26] for the correlation functional, optimized norm-conserving Vanderbilt pseudopotenetials [27] with spin-orbit coupling for core electrons, 10×10×2 for k-sampling integration and 40 Ry for energy cutoff. Further details are in Supplemental Material [14].

## III. RESULTS

### A. Standing-wave excited photoemission and rocking curves

A SW with its iso-intensity planes parallel to the diffracting planes is created by the interference between the incident ($k_0$) and diffracted ($k_{002}$) waves [9], as illustrated in Fig. 1. Figure 1(a) shows the Bi2212 crystallographic structure of the top half unit cell, and its cleavage plane primarily occurs in between the BiO layers due to weak van der Waals bonds [1]. The Bi, Sr, Cu, and Ca cations in these layers are well separated along the c-axis direction, making Bi2212 an ideal candidate for applying the SW technique to derive layer-resolved information. From SW theory based on dynamical x-ray diffraction [8,9], the phase difference between the incident and diffracted wave fields changes by $\pi$ when the incidence angle moves from below to above the Bragg condition, thus scanning the SW by $d_{002}/2$ with respect to the (002) planes, as illustrated in Figs. 1(b) and 1(c). The detailed theoretical SW modeling, including consideration of both x-ray and electron attenuation with depth, is discussed in Supplemental Material S3 [14]. Depending on the locations of the atoms with respect to the scanned SW, the incidence-angle dependence of the core-level photoelectron intensities, which we define as core-level rocking curves (RCs), will show distinct modulations as to both shape and magnitude.

Figures 2 and 3 illustrate the layer-dependent results for core-level intensities using SW excitation. The photoelectron spectra of Ca 2p, Sr 3d, Cu 3p, and Bi $4f_{7/2}$ at an off-Bragg angle (23.2°) are shown in Figs. 2(a)-(d). Such photoelectron spectra were collected by varying the incidence angle between 24° and 27.5°, yielding the five distinct core-level RCs in Figs. 2(e)-(i). These RCs are normalized to 1 at off-Bragg positions and have been simulated by SW theory (red curves in Figs. 2(e)-(i)). Both members of the spin-orbit split Ca 2p spectrum (Fig. 2(a)) exhibit two components, with low-binding-energy (LBE) peaks at 344.6 and 348.1 eV and high-binding-energy (HBE) features at 345.9 and 349.4 eV. These have been observed in previous XPS studies, with varying relative intensities, depending on the sample synthesis procedure [6,28,29,30]. The RCs of the Ca 2p(LBE) in Fig. 2(e) and Ca 2p(HBE) in Fig. 2(f) show different shapes and relative intensity modulations, with LBE exhibiting higher modulation ~8%, as compared to ~5% for HBE. The experimental Ca 2p(HBE) RC is, within statistical noise, also identical to the Sr 3d RC (Fig. 2(g)), including the amplitude of modulation, suggesting that the depths of Ca(HBE) and Sr atoms are essentially identical. By fitting these two experimental Ca 2p RCs using Eqs. (S2) and (S3), we are able to derive the values of coherent position ($P_{HQ}$), which determines the shape of RCs and provides the averaged locations of Ca(HBE) and Ca(LBE) atoms. Detailed discussions regarding the parameters on fitting the experimental RCs and a summary table S1 can be found in Supplemental Material S3 [14]. Thus, we can unambiguously conclude that the Ca(LBE) atoms are located in the Ca layer, while the Ca(HBE) atoms are located in the SrO layer, implying that a significant fraction of Ca atoms occupy the Sr sites during synthesis.

A more quantitative analysis was made by considering the peak intensity ratio, I(HBE)/[I(HBE)+I(LBE)] away from the Bragg reflection, which is ~0.2 with respect to the ideal amount of Ca; this indicates that an excess of ~10% Ca intermixing with each of the two adjacent SrO layers. Previous work on the degree of Sr-Ca intermixing is controversial [6,28,29,30], with some studies suggesting pronounced Sr-Ca intermixing in both Ca and SrO layers [28] and some claiming low intermixing but with strong dependence on sample preparation [29,30]. For our sample, both the observation of only one component in the Sr 3d spectrum and its RC show that the Sr atoms are located in a single layer without intermixing. Note that although the chemical composition of Bi2212 here is referred to $Bi_2Sr_2CaCu_2O_{8+\delta}$, the actual thermodynamically stable composition can be deficient in Sr and Ca while being Bi rich. E.g. Mitzi *et al.* found that the stable composition, by normalizing Cu to be 2, is $Bi_{2.03}Sr_{1.87}Ca_{0.85}Cu_2O_{8+\delta}$ and the precise numbers in fact vary from sample to sample [31]. In our work, from two successful SW-XPS measurements on cleaved Bi2212 samples, the quantity of intermixing shows no noticeable difference; however, these two samples come from the same large crystal and should exhibit similar stoichiometry. The quantity of intermixing can vary with different crystal preparations [31], a possible subject of future study with standing-wave excitation.

The Cu 3p and Bi 4f spectra in Figs. 2(c) and 2(d) also show single components, although Cu 3p is broad, as seen previously [7], and their very different RCs (Figs. 2(h)-(i)) demonstrate that the Cu and Bi atoms are uniquely located in their own layers. Note that the *shape* of the Sr 3d and Cu 3p RCs are close, which is not surprising in view of the location of Sr atoms on either side of the Cu atoms, but the Cu 3p RC shows a stronger intensity modulation due to the lack of SW phase averaging over the two Sr layers in the half unit cell. All of these conclusions are supported by the excellent agreement between experiment and SW modeling in Figs. 2(e)-(i).

We now consider the O 1s spectrum in Fig. 3(a), which is thought from prior XPS work to exhibit three components contributed from the different atomic layers [32,33]. Through modeling the O1s RCs with SW theory, the locations of these components were determined. The O 1s(P1) RC (Fig. 3(b)) shows that these oxygen atoms are located in the BiO layer. The O 1s(P2) RC in Fig. 3(c) has a similar shape but weaker intensity modulation compared to the Cu 3p RC, suggesting that, in the first CuO layer, the O(P2) atoms is ~0.9±0.5 Å higher than Cu atoms. O 1s(P3) RC (Fig. 3(d)) is slightly out of phase with respect to the Sr 3d RC, suggesting that, in the first SrO layer, the P3 oxygen atoms are ~1.5±0.5 Å higher relative to Sr atoms. These SW-determined locations of the oxygen atoms are in good agreement with prior transmission electron microscopy and x-ray diffraction results [24,34]. Looking ahead, future SW photoemission studies of Bi2212 or other cuprates, with higher reflectivities and better statistics, and with various oxygen dopant levels, should be able to determine the O stoichiometries in each layer and thus answer the question of where the additional oxygen dopant atoms reside.

In addition, these SW results provide unique insight into the chemical/electronic disorder along the c-axis in cuprates. In a broader perspective, several cuprate studies have demonstrated the interesting out-of-plane electronic properties. For example, YBCO exhibits a three-dimensional charge ordering at high magnetic fields [35,36]. The c-axis resistivity has been used to reveal information on the pseudogap phase [37] and magneto-transport [38] as well as soft x-ray ARPES [39] about the Fermi surface warping. Coming back to our work, the chemical disorder along the c-axis might prevent three-dimensional charge order in Bi2212 and also give rise to a larger scattering in c-axis resistivity experiments. Future experiments exploring these chemical/electronic effects in more detail should be very interesting.

**B. Atomic-layer-resolved electronic structure**

In order to resolve the individual atomic layer contributions to the Bi2212 valence band (VB), we measured the VB RCs, which is an intensity map $I_{VB}(E_b, \theta_{inc})$, with binding energy ($E_b$) and incidence angle ($\theta_{inc}$), over an angle scan. The VB RCs can be written as a superposition of the experimentally layer-projected and cross-section weighted DOSs in the different layers, $D_i(E_b) \approx \sum_{i,Qnl} \frac{d\sigma_{Qnl}}{d\Omega} \rho_{Qnl}(E_b)$, multiplied by layer-dependent normalized core-level RCs $\bar{I}_{Qn'\ell'}(\theta_{inc})$, as discussed previously [11,12,13,40,41,42,43], see further details in Supplemental Material S4 [14]. That is,

$$I_{VB}(E_b,\theta_{inc}) = \sum_{Qn\ell} D_{Qn\ell}(E_b) \bar{I}_{Qn'\ell'}(\theta_{inc}) \quad . \tag{1}$$

Here $Qn\ell$ denotes a valence level $n\ell$ in the atom $Q$, and $Qn'\ell'$ a core level in the same atom, and $\bar{I}_{Qn'\ell'}(\theta_{inc})$ is the RC for $Qn'\ell'$, normalized to unity away from the Bragg reflection. For our case, $Qn'\ell'$ = Cu 3p, Sr 3d, and Bi $4f_{7/2}$.

The main contributions from the atomic orbitals in the layer-projected DOSs based on strength of hybridization and photoelectric cross sections at our excitation energy are Cu 3d in $CuO_2$, Sr 4p in SrO, and Bi 5d in BiO (see Supplemental Material S5 [14]). The Ca 4s orbitals in the Ca layer are negligible, as discussed in prior work using density functional theory (DFT) calculations in the local-density approximation (LDA) [44,45].

In both the raw data for $I_{VB}(E_b, \theta_{inc})$ of Fig. 4(a) and in a more pronounced way in its second derivative along the axis of incidence angle in Fig. 4(b), there are intensity modulations with BE associated with the different layer contributions to the intensity; these are particularly clear along the Bragg angle at ~25.7°. The VB RCs for each BE have been fitted to a linear combination of these three core-level RCs by a least-squares method, and the resultant fitting coefficients correspond to the layer-resolved cross-section weighted DOSs $D_i(E_b)$.

Fig. 4(c) shows these experimental layer-projected cross-section-weighted DOSs (dots) in comparison to DFT calculations (curves) incorporating the supermodulation displacements known to exist in Bi2212 [14], in particular with a twofold enlargement of the unit cell size in the x-y plane (see Supplemental Material S6 [14]). The $D_{CuO_2}(E_b)$ and $D_{BiO}(E_b)$ show good agreement with the DFT results including supermodulation. The $D_{SrO}(E_b)$ is however considerably broader than the DFT results. One obvious source of uncertainty and broadening has been mentioned before: the Sr 3d RC has similar shape compared to the Cu 3p RC, meaning that the deconvolution procedure will inherently mix some intensity from $CuO_2$. Beyond this, the broadening of $D_{SrO}(E_b)$ is certainly associated with the significant Ca-Sr intermixing, causing more disorder and scattering. Further electronic structure calculations, e.g. within the coherent potential approximation (CPA), would help to test this hypothesis.

## IV. DISCUSSION

To assist in understanding the influence of the supermodulated atomic displacements on the electronic structure, the layer-projected DOSs with and without supermodulation are plotted together in Fig. 4(d). The introduction of supermodulation makes no significant difference in the CuO$_2$ DOS, but for both cases, a shift to higher BE by 1.3 eV is necessary to reach agreement with the experimental $D_{CuO_2}(E_b)$. For the SrO and BiO DOSs, no BE shift is needed for the supermodulation results, whereas without supermodulation, one needs to shift the curves to higher BE by 0.9 eV and to lower BE by 1.3 eV to best match experiment, respectively. Adding supermodulation for the SrO DOS further produces an additional peak at ~5 eV that much better matches the experimental $D_{SrO}(E_b)$. For the BiO plane, the disappearance of a peak at ~6.5 eV in the theoretical DOS with supermodulation structure is again in better agreement with the experimental $D_{BiO}(E_b)$. In summary, taking supermodulation into consideration leads to a change of the SrO and BiO DOSs, improving the agreement with the layer-projected DOSs. This indicates the strong influence of the supermodulation structure on the electronic structure of Bi2212.

We have noted that an energy shift of the theoretical CuO$_2$ DOS to 1.3 eV higher BE has been necessary to match the $D_{CuO_2}(E_b)$. Such shifts of the VB energies in photoemission relative to LDA calculations have been widely reported in the cuprates (e.g. LSCO [46], YBCO [46], and Bi2212 [47]). In undoped cuprates the Coulomb interaction between the Cu 3$d$ electrons in cuprates, which is not treated fully in simple DFT calculations, can lead to the opening of a Mott-Hubbard gap, with a bound-state energy shift to higher BE and a lower DOS in the vicinity of the Fermi energy (E$_F$) [46,47]. These features in the cuprate VBs are spectroscopic evidence of strong correlation effects, and more detailed discussions can be found elsewhere [48]. In metallic cuprates, where the Cu 3d states hybridized with O 2p dates mostly lie 2-6 eV below the Fermi level, the main effect relevant to the current data is an increase in binding energy. The $D_{CuO_2}(E_b)$ thus shows that the energy shift that can be attributed to this electron correlation effect is 1.3 eV.

In order to visualize in more detail the layered-resolved electronic structure of Bi2212, the band structures near the $E_F$ region are shown in Fig. 5. The full band structure without supermodulation is shown in Fig. 5(a), and in Fig. 5(b) with supermodulation. The band structure in Fig. 5(a) is in good agreement with prior work [44,45]. By comparing these two figures, one clearly sees a splitting of the bands (at ~-2 eV and at 0.5 to 2 eV BE) that results from including supermodulation. For the more realistic band structure with supermodulation we now show the layer-projected band structures in Figs. 5(c)-(e) in a blue-gray scale to indicate relative amplitude. From these results alone, one would conclude that, around the M point, the Fermi surface of Bi2212 is governed by the $CuO_2$ bands, but that there is also a strong contribution from BiO bands. Although our DFT results show the existence of BiO bands near the M point (Fig. 5(e)) and some BiO state intensity extending below $E_F$ (Fig. 4(c)), our experimental results in the same figure lack that spectral feature, in agreement with previous photoemission studies [1,6]. For example, to resolve this disagreement between the DFT results and photoemission, Lin *et al.* [49] proposed that, with increasing oxygen doping in the BiO layer, the BiO band shifts above $E_F$ at the M point, which also is consistent with a scanning tunneling microscopic and spectroscopic study [50]. The excess oxygen atoms are believed to be responsible for the δ in the common designation $Bi_2Sr_2CaCu_2O_{8+\delta}$.

## V. CONCLUSIONS

In summary, we have carried out soft x-ray SW photoemission study of Bi2212 and derived the depth distribution of atoms within one unit cell, in particular, a 10% Ca-Sr intermixing and the three types of oxygen atoms bonding to different cations. In addition, we have successfully decomposed the electronic structure of Bi2212 into atomic-layer-specific, matrix-element-weighted DOSs. These atomic-layer-resolved DOSs show good agreement with DFT calculations in most respects, provided we incorporate the known supermodulation structure in Bi2212. Our results for the layer-resolved electronic structure are found to be strongly influenced by the supermodulation, Ca-Sr intermixing, and the Cu 3d-3d Coulomb interaction, further clarifying the complexity of this prototypical cuprate. Future measurements of this type for other cuprates should yield equally unique information, such as providing insights on how the $T_c$ increases while stacking more $CuO_2$ layers from bi-layered Bi2212 to tri-layered $Bi_2Sr_2Ca_2Cu_3O_{10+\delta}$ (Bi2223). Bragg-reflection SW photoemission is thus very promising for the study of quasi-two-dimensional materials with large-c lattice parameters.

**ACKNOWLEDGMENTS**

The authors would like to thank Jonathan Denlinger and Simon Moser for their comments concerning Bi2212 sample preparation. We thank synchrotron SOLEIL (via Proposal No. 20161205) for access to Beamline CASSIOPEE that contributed to the results presented here. This work was supported by the US Department of Energy under Contract No. DE-AC02-05CH11231 (Advanced Light Source), and by DOE Contract No. DE-SC0014697 through the University of California Davis (salary for C.-T.K, S.-C.L. and C.S.F.). S.-T. Pi was supported by DOE grant DE-NA0002908. C.S.F. has also been supported for salary by the Director, Office of Science, Office of Basic Energy Sciences (BSE), Materials Sciences and Engineering (MSE) Division, of the U.S. Department of Energy under Contract No. DE-AC02-05CH11231, through the Laboratory Directed Research and Development Program of Lawrence Berkeley National Laboratory, through a DOE BES MSE grant at the University of California Davis from the X-Ray Scattering Program under Contract DE-SC0014697, through the APTCOM Project, "Laboratoire d'Excellence Physics Atom Light Matter" (LabEx PALM) overseen by the French National Research Agency (ANR) as part of the "Investissements d'Avenir" program, and from the Jűlich Research Center, Peter Grűnberg Institute, PGI-6. Support for W.E.P. was provided by DOE grant DE-FG02-04ER46111. This research used resources of the National Energy Research Scientific Computing Center (NERSC), a DOE Office of Science User Facility supported by the Office of Science of the U.S. Department of Energy under Contract No. DE-AC02-05CH11231.

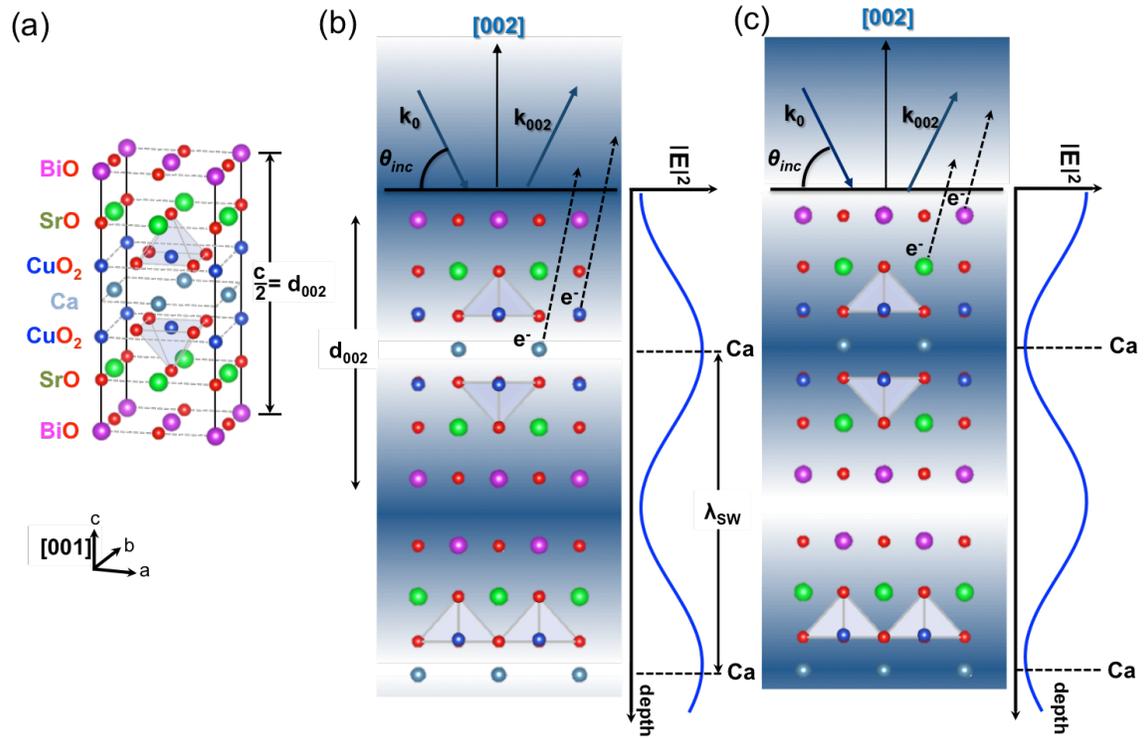

FIG. 1. Spatial relationship between the $Bi_2Sr_2CaCu_2O_{8+\delta}$ (Bi2212) crystal planes and the (002) Bragg-reflection x-ray standing wave (SW). (a) Top half of the unit cell of Bi2212. (b) Schematic of the experimental geometry and the SW generated by the Bi2212 (002) reflection, with wavelength = $\lambda_{SW}$ = c/2 = $d_{002}$. The incident and diffracted waves (associated with wave vectors $k_0$ and $k_{002}$) interfere to produce the SW. The photon energy was 930.3 eV, with a corresponding Bragg angle of about 25.7°. (c) By increasing the incidence angle around the (002) Bragg reflection, the SW can be shifted by $d_{002}/2$. The maximum of the SW electric field intensity can thus be shifted continuously from the Ca plane (b) to the BiO plane (c).

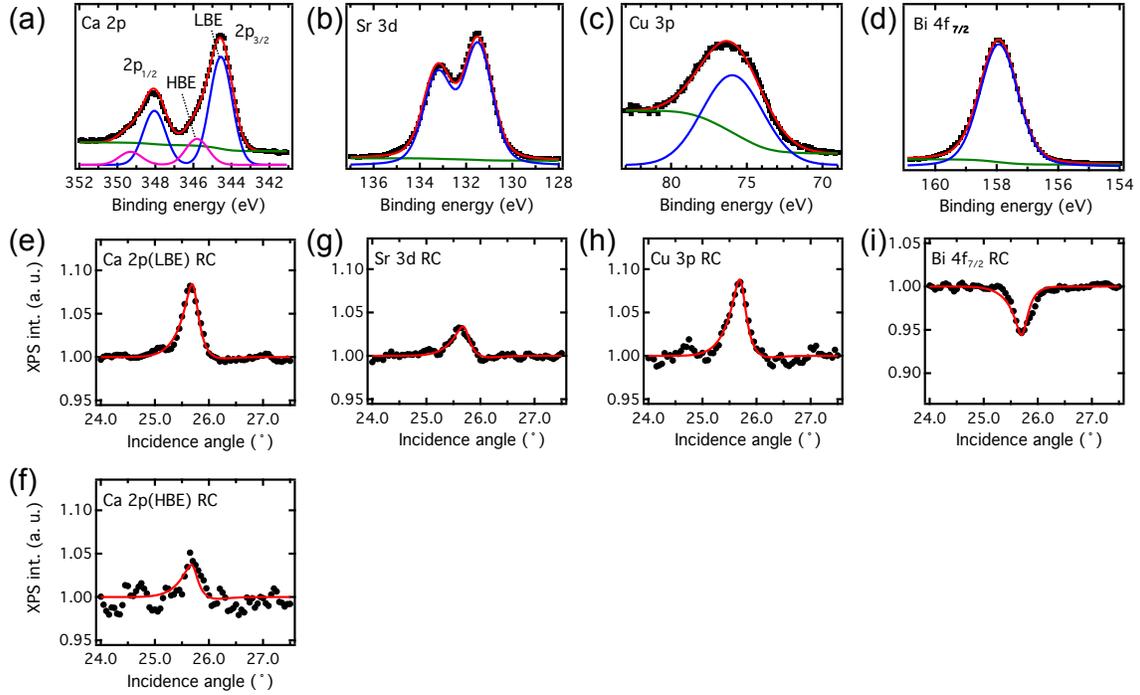

FIG. 2. Bragg-reflection standing-wave x-ray photoemission from the cations of Bi2212 at $h\nu$ = 930.3 eV. Core-level spectra of (a) Ca 2p, (b) Sr 3d, (c) Cu 3p, and (d) Bi $4f_{7/2}$ at an off-Bragg incidence angle. The core-level peak intensities are derived by fitting with a Voigt line shape (in blue and magenta) and a Shirley background (in green). The corresponding experimental rocking curves (RCs) of core-level intensities are also plotted in (e)-(i) (black dots) and compared with SW theory (red curves). In (a) Ca 2p is found to have high-binding-energy (HBE) and low-binding-energy (LBE) components, which shows different RC behavior as to shape and fractional modulation. The RCs of Ca 2p HBE (in (f)) and Sr 3d (in (g)) are found to be identical within experimental error, indicating that Ca atoms occupy the Sr sites in the SrO layer.

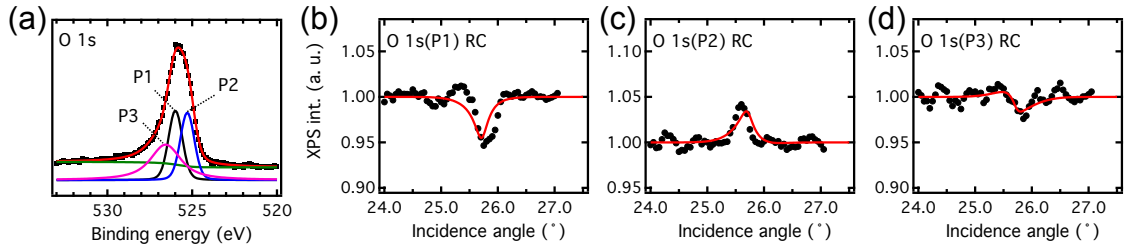

FIG. 3. Bragg-reflection standing-wave x-ray photoemission from the oxygen atoms of Bi2212. (a) Core-level spectrum of O 1s at an off-Bragg incidence angle. The O1s spectrum is known to contain three components (P1, P2, and P3). The RCs of (b) P1, (c) P2, and (d) P3 exhibit distinct shape and intensity modulations; these can be assigned through SW analysis (red curves) to different layers. The O(P1) atoms are located in the BiO layer. The O(P2) atoms are in the $CuO_2$ layer with a vertical offset of ~0.9±0.5 Å to the Cu atoms. The O(P3) atoms are in the SrO layer with a vertical offset of ~1.5±0.5 Å.

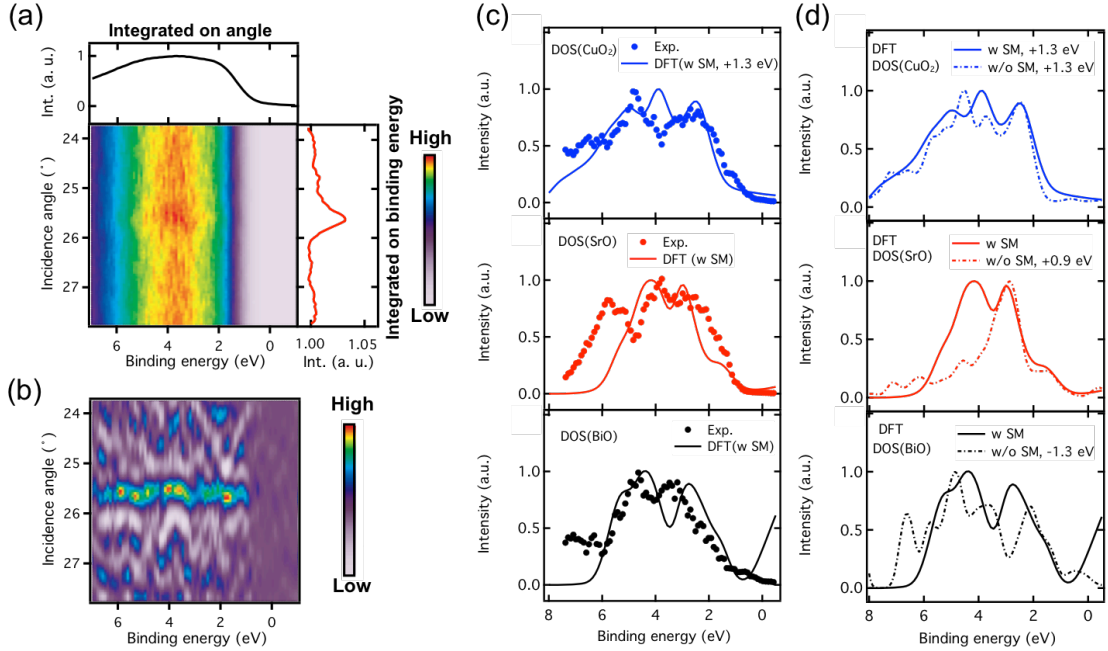

FIG. 4. Standing-wave valence-band (VB) spectra and atomic-layer-resolved, cross-section-weighted densities of states (DOSs) together with DFT calculations. (a) VB intensity map for different incidence angles, with the color scale corresponding to the photoemission intensity, and top and right curves representing integrals over incidence angle and binding energy, respectively. (b) As (a), but for the second-derivative of intensity. The Bragg–reflection maximum is at ~25.7°, and the VB intensities exhibit modulations that are associated with the variable layer-specific contributions along the binding energy axis; this is particularly evident in the second derivative plot. (c) The layer-projected DOSs of $CuO_2$ (blue dots), SrO (red dots), and BiO (black dots) and the corresponding layer-projected DOSs from DFT calculations including supermodulation in the crystal structure (solid line). (d) The comparison of DFT results with supermodulation (w SM) structures (solid line) and without supermodulation (w/o SM) structures are shown (dot-dashed line). Various energy shifts have been applied to theory in (c) and (d) to yield the best agreement with experiment; with supermodulation, only the $CuO_2$ DOS requires a shift.

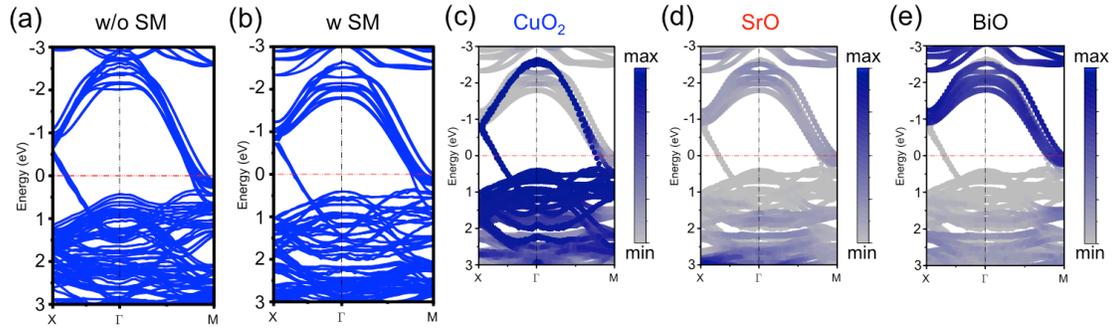

FIG. 5. Band structure in the high-symmetry directions near the $E_F$. Full band structure for Bi2212 without (a) and with (b) supermodulation effect in the crystal structure. The layer-projected (c) $CuO_2$, (d) SrO, and (e) BiO band structures of (b) (with supermodulation). The blue-to-grey color scale indicates strong-to-weak contribution from a given layer. Note that no energy shift is applied to theory here.

# Supplemental Material

# Atomic-layer-resolved composition and electronic structure of the cuprate $Bi_2Sr_2CaCu_2O_{8+\delta}$ from soft x-ray standing-wave photoemission

Cheng-Tai Kuo,[1,2,*] Shih-Chieh Lin,[1,2] Giuseppina Conti,[1,2] Shu-Ting Pi,[1] Luca Moreschini,[3] Aaron Bostwick,[3] Julia Meyer-Ilse,[2] Eric Gullikson,[2] Jeffrey B. Kortright,[2] Slavomir Nemšák,[3,4] Julien E. Rault,[5] Patrick Le Fèvre,[5] François Bertran,[5] Andrés F. Santander-Syro,[6] Ivan A.Vartanyants,[7,8] Warren E. Pickett,[1] Romuald Saint-Martin,[9] Amina Taleb,[5] and Charles S. Fadley[1,2,†]

**Affiliations**

[1]Department of Physics, University of California Davis, Davis, California 95616, USA

[2]Materials Sciences Division, Lawrence Berkeley National Laboratory, Berkeley, California 94720, USA

[3]Advanced Light Source, Lawrence Berkeley National Laboratory, Berkeley, California 94720, USA

[4]Peter Grünberg Institut PGI-6, Research Center Jülich, 52425 Jülich, Germany

[5]Synchrotron SOLEIL, L'Orme des Merisiers, Saint-Aubin-BP48, 91192 Gif-sur-Yvette, France

[6]CSNSM, Université Paris-Sud, CNRS/IN2P3, Université Paris-Saclay, 91405 Orsay Cedex, France

[7]Deutsches Elektronen-Synchrotron DESY, Notkestraße 85, D-22607 Hamburg, Germany

[8]National Research Nuclear University MEPhI (Moscow Engineering Physics Institute), Kashirskoe Shosse 31, Moscow, 115409, Russia

[9] SP2M-ICMMO-UMR-CNRS 8182, Université Paris-Sud, Université Paris-Saclay, 91405 Orsay Cedex, France

[*] Corresponding author: chengtaikuo@lbl.gov
[†] Corresponding author: fadley@lbl.gov

## S1. Sample preparation and characterization

Bi$_2$Sr$_2$CaCu$_2$O$_{8+\delta}$ (Bi2212) single crystal growth was performed in an optical floating zone furnace (Crystal System Incorporation, Japan) equipped with four 300W lamps installed as infrared radiation sources. The sintered feed rods used for the crystal growth, prepared by the conventional solid-state reaction, were pre-melted in air with a mirror scanning velocity of 15mm/h by traveling the upper and lower shaft to densify the feed rod and to avoid the emergence of bubbles during the crystal growth. A previously grown Bi2212 crystal ingot was used as a seed rod. The feed rod and the growing crystal were rotated at 15 rpm and 20 rpm, respectively, in opposite directions to ensure efficient mixing and uniform temperature distribution in the molten zone. Different growth rates (0.5 to 2.5 mm/h) and atmosphere pressure (1, 2 and 3 bars) were applied but the highest-quality single crystals were obtained for a slow growth rate of 0.6mm/h and an oxygen pressure p(O$_2$) = 2 bar. X-ray diffraction measurements were performed in order to rule-out the possible presence of a secondary phase. The critical temperature, which is determined from the onset of the SQUID diamagnetic transition, is ~93K.

Before the photoemission measurements, the Bi2212 single crystal was cooled down to ~77K by liquid N$_2$ below its superconducting transition temperature and *in situ* cleaved in the analysis chamber. Angle-resolved photoemission spectroscopy (ARPES) at hv = 110 eV was used for initial sample alignment. Figure S1(a) shows the Fermi surface of Bi2212 measured by ARPES at 110 eV and it shows the typical patterns as the prior studies [S1,S2] The measurement was performed to confirm a well-ordered region of the cleaved surface by scanning in x and y and to precisely align the crystal to the Bragg reflection condition; this was repeated after our standing-wave excited photoemission (SW-XPS) measurements to verify surface stability. The SW-XPS was measured at 930.3 eV, for which we were unable to observe any momentum resolution, likely due to some combination of the supermodulation in the crystal and the momentum-smearing effects of phonons. The details of choosing a suitable SW excitation energy are discussed in the next section. The beam size was 100 μm x 200 μm. The binding energy was calibrated using a gold reference sample. The estimated experimental resolution for the SW measurements is 0.6 eV. A survey spectrum (Fig. S1(b)) of the cleaved Bi2212 surface shows core levels from all the expected elements present, with no indication of a surface contaminant signal from C 1s; surveys before and after our measurements yielded the same conclusion.

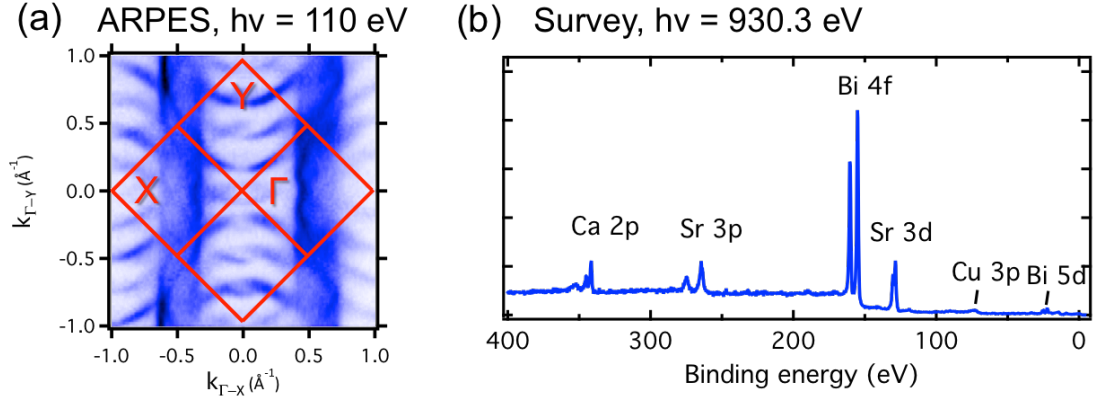

FIG. S1. Photoemission characterization of cleaved Bi2212. (a) ARPES Fermi surface obtained at hv = 110 eV. (b) Survey spectrum obtained at our SW energy of hv =930.3 eV.

**S2. Cu $L_3$ edge resonant effect on absorption, reflectivity and SW strength**

According to SW theory based on dynamical x-ray diffraction [S3,S4,S5], the strength of the SW is proportional to the square root of the reflectivity, R, with overall amplitude of modulation of ~$2(R)^{1/2}$. In order to maximize the reflectivity, and thus the SW effect, the excitation energy was varied through the strong Cu $L_3$ absorption edge (Fig. S2(a)), as demonstrated recently for multilayer oxide heterostructures [S6]. Reflectivity was measured at Beamline 6.3.2 of the Advanced Light Source. Figure S2(b) plots the reflectivity as a function of the excitation energy (blue curve), and it shows that the maximum of reflectivity is at 930.3 eV, which is ~1.6 eV below the $L_3$ x-ray absorption (XAS) peak (black curve), also shown in this panel. The reflectivity for two energies near the resonance (cuts A and B in Fig. S2(a)) has also been analyzed using SW theory (green and red curves in Fig. S2(c)), and they exhibit excellent agreement with experiment (green and red dots in Fig. S2(c)), including the marked increase in reflectivity on going just below the absorption maximum (cf. also Fig. S2(b)). The angular scan of reflectivity at 930.3 eV (cut A in Figs. S2(a) and S2(c)) has a maximum of ~2.6x10$^{-4}$, leading to a maximum of standing wave modulation of ~ $2\sqrt{R}$ ~ 3% (see further details for SW modeling below). Note that the RC modulations in Figs. 2(e)-(i) in the article are in the 5-10% range, roughly consistent with this estimate, especially since we could not cleave the sample surface for the reflectivity measurement. For our reflectivity measurement (BL 6.3.2), the size of the Bi2122 crystal needs to be much larger than what it is needed for the SW photoemission measurement (BL 7.0.2). Due to the limitation of available large-size Bi2212 crystals with flat surface, the rough sample surface for the reflectivity measurement leads to the lower reflectivity and the lower $2\sqrt{R}$ estimate (~ 3%) with respect to experimental SW modulation (5-10%) measured by SW photoemission. For comparison, the maximum of reflectivity at the XAS peak (cut B in Figs. S2(a) and S2(c)) shows a decrease to ~1x10$^{-4}$ that would reduce the SW modulation by about one half. Therefore, all our SW-XPS measurements were carried out at hv = 930.3 eV.

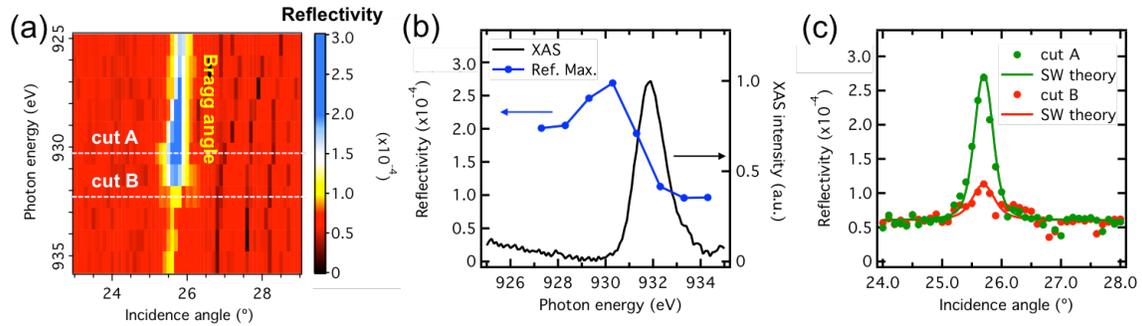

FIG. S2. Resonant effects on the (002) Bragg reflection in Bi2212 near the Cu L$_3$ edge. (a) Reflectivity as a function of incidence angle and photon energy near the Cu L$_3$ edge. (b) Comparison of the maximum reflectivity (Ref. Max., left axis, blue line) as a function of photon energy and the x-ray absorption spectrum (XAS, right axis, black line). (c) Reflectivity as a function of angle for two photon energies (cuts A and B in (a)). The points are angular scans of reflectivity measured around the Bragg angle at hv = 930.3 (cut A in (a)) and 932.3 eV (cut B in (a)). The experimental data (points) are compared with the SW theory (curves).

## S3. Core-level rocking curves: SW modeling based on dynamical diffraction

The normalized core-level rocking curves (RCs) in Figs. 2,3 of the main text have been analyzed using SW theory based on dynamical x-ray diffraction, as applied in particular to photoemission [S3,S4,S5]. The SW intensity for a given H = (*hkl*) reflection of a given element Q from a given depth $z_i$ below the surface is

$$I_{SW,HQ}(z_i,\theta_{inc}) = I_0 \exp^{-\frac{z_i}{\Lambda_x^{eff}\sin\theta_{inc}}} \left[ 1 + R(\theta_{inc}) + 2C\sqrt{R(\theta_{inc})} f_{HQ} \cos(\varphi_H(\theta_{inc}) - 2\pi P_{HQ}(z_i)) \right], \quad \text{(S1)}$$

where $I_0$ is the incident intensity, $\Lambda_x^{eff}$ is the effective attenuation length of x-ray due to both absorption and diffraction, $R$ is the reflectivity at a given incidence angle $\theta_{inc}$, measured with respect to the surface, $C=2\cos(2\theta_B)$ is the polarization factor for π-polarization, $\theta_B$ is the Bragg angle for the *H* reflection, $f_{HQ}$ is the coherent fraction of atoms of type Q for the *H* reflection, $\varphi_H$ is the phase difference between the incident and diffracted waves, and $P_{HQ}$ is the coherent position of atom *Q* for the *H* reflection at depth $z_i$. For our Bi2212 sample, direct calculations reveal that the absorption length is ~75 nm and the extinction length due to diffraction is 1775 nm; thus $\Lambda_x^{eff} \approx 71$ nm is much greater than the inelastic mean free path for electrons, which is ~1-1.5 nm, and so we can set the exponential to unity in Eq. (S1).

An effectively angle-integrated photoemission intensity for a give *H* reflection, emission from the n$\ell$ level of a given element *Q*, which is the core-level RC, can then be calculated as

$$I_{Qn\ell}(\theta_{inc}) = \rho_Q \frac{d\sigma_{Qn\ell}}{d\Omega} \left[ 1 + R(\theta_{inc}) + \sum_i^N \frac{e^{\frac{-z_i}{\Lambda_e \sin\theta_e}}}{I_A} \times 2\cos(2\theta_B)\sqrt{R(\theta_{inc})} f_{HQ} \cos(\varphi_H(\theta_{inc}) - 2\pi P_{HQ}(z_i)) \right], \quad \text{(S2)}$$

where $\rho_Q$ is the density of atom *Q* and $d\sigma_{Qn\ell}/d\Omega$ is the differential photoelectric cross section of level $Qn\ell$. The exponential allows for photoelectron attenuation due to inelastic scattering, where $\Lambda_e$ is the electron inelastic mean-free paths and $\theta_e$ is the electron emission angle with respect to the sample surface. $I_A$ is a normalization factor for photoelectron intensities that represents the sum over all layers in the absence of the SW effect: $I_A = \sum_j^N e^{\frac{-z_j}{\Lambda_e \sin\theta_e}}$. In each core-level RC, the intensity is finally normalized to 1 at off-Bragg position as the atom density and differential don't affect its phase and modulation of intensity, thus providing an element-specific measure of the fractional modulation due to the SW at a given atomic type.

Figure S3 shows the geometry of our SW measurement, and illustrates the definition of incidence and emission angles: e.g., for our geometry, $\theta_e = \theta_{inc} + 54°$. $Qn\ell$ for our case thus involves layers containing Cu 3p, Sr 3d, Bi $4f_{7/2}$, two types of Ca 2p, and three of O 1s (denoted P1, P2, and P3 in the text). In Eq. (S2) the summation was taken with N over 3 unit cells, which translates through the electron inelastic mean free path to including 99.6% of the photoelectron intensity. The photoelectrons were collected in a partially angle-integrated manner, as the analyzer collected photoelectrons from the +/- 6 degrees of the analyzer lens axis, and emission was near the (001) sample normal.

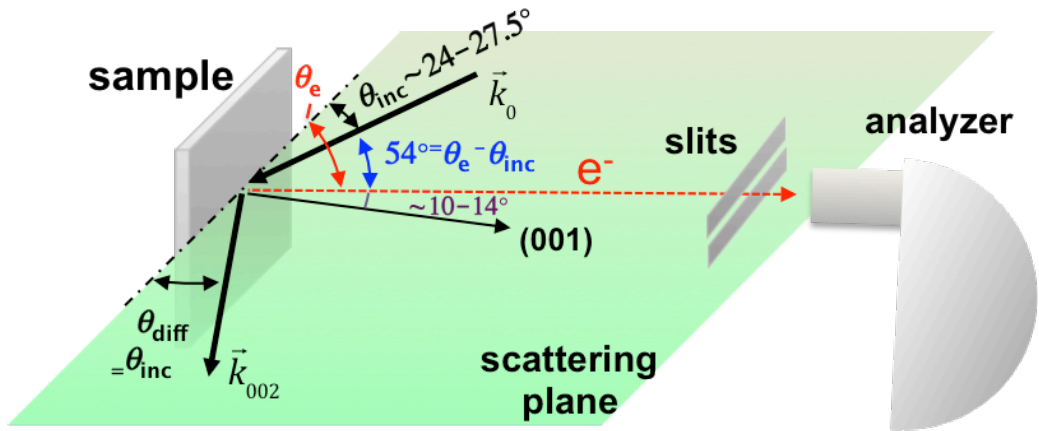

FIG. S3. Geometry of SW-XPS measurement. The incidence angle between the incident light with wave vector and the sample surface varies from 24 to 27.5°. The experimental geometry fixes the angle between the incident light and the outgoing photoelectrons at 54°.

The reflectivity of an Bi2212 supermodulated crystal, $R(\theta_{inc})$ and phase $\varphi_H(\theta_{inc})$, including dynamical diffraction effects, were calculated from Eq. (S3) (Ref. [S3]), which contains three fitting parameters in the factors at right:

$$R(\theta_{inc}) = \left|\frac{E_H}{E_0}\right|^2 = \left|\eta \pm (\eta^2-1)^{\frac{1}{2}}\right|^2 \left|\frac{F_H}{F_{\bar{H}}}\right|^2 \times \frac{1}{\sigma\sqrt{2\pi}} e^{-\frac{\Delta\theta^2}{2\sigma^2}} \times e^{-W}. \tag{S3}$$

Here, $E_H$ and $E_0$ are the complex electric field amplitudes of the incident and diffracted waves, with $H = (002)$ reflection, $\eta$ is defined below and is a normalized angle parameter dependent on the deviation of the angle $\theta_{inc}$ from the Bragg angle $\theta_B$, $F_H$ and $F_{\bar{H}}$ are the structure factors for $H$ and $\bar{H}$ reflections, σ is the Gaussian width of the reflectivity curve, incorporating e.g. the x-ray beam divergence, and W is the Debye-Waller factor = $\exp(-H^2 \langle u^2\rangle /3)$, where H is the scattering vector, and $\langle u^2 \rangle$ is the mean-square atomic displacement. Thus, σ determines the width of the reflectivity curve and the RCs, and $W$ determines the intensity of reflectivity and the RC modulation. The first factor before the multiplication sign in $R(\theta_{inc})$ is the reflectivity from an Bi2212 supermodulated crystal. Here, we considered the supermodulation in a twofold enlargement of the unit cell [S7]. Detailed information regarding the supermodulation structure is discussed in Ref. [S7] and Section 6. The next factor allows for mosaicity in the crystal and x-ray beam divergence and the fourth factor for vibrational motion and static distortions of atomic positions from the ideal structure, as e.g. the supermodulation of atomic positions in the crystal structure. Adding the two factors to reduce the intensity and broaden the peak width has been widely used in the x-ray SW fluorescence studies of thin films [S8,S9], in our case, they allow for a combination of vibrational motion and static distortions of atomic positions from the ideal structure, the presence of any defects, and the supermodulation of atomic positions in the crystal structure. Note that the SW phase is not affected by these two factors. The normalized angle parameter $\eta$ is defined as:

$$\eta = \frac{-\Delta\theta \sin 2\theta_B + \frac{r_e \lambda_x^2}{\pi V} F_0}{\frac{r_e \lambda_x^2}{\pi V} \sqrt{F_H F_{\bar{H}}}}, \tag{S4}$$

where $\Delta\theta = \theta_{inc} - \theta_B$, $r_e$ is the classical electron radius, $\lambda_x$ is the wavelength of the x-ray, $V$ is the total volume of the unit cell, and $F_0$ is the structure factor for (000) reflection.

The σ and $W$ values determined from a combined analysis of the reflectivity and SW photoemission data are 0.28±0.02° and 2.95±0.04, with the large value for the second factor probably being due to the supermodulation and residual surface roughness after the cleave. Compared to our model of 2-fold larger unit cell, the actual Bi2212 crystals have 5-fold larger unit cell [S10], which can significantly reduce the reflectivity and lead to an extreme W value of ~3. Note that the low reflectivity is due to the several factors, including the intrinsic supermodulation of the Bi2212 crystal, and not any sort of damage due to the *in situ* cleavage. After fitting all of the core-level RCs, we derive $f_{HQ}$ and $P_{HQ}$ values, as well as the absolute positions ($z_Q$) of the first contributing layers, and these are shown in Supplemental Table S1. These numbers confirm our assignment of Ca(HBE) to the SrO layer, a displacement of the O atoms from the Cu atoms in the CuO layer, and a displacement of the O atoms in the Sr layer, as discussed in the main text. These displacements are consistent with prior transmission electron microscopy and x-ray diffraction results for Bi2212 [S7,S11]. The quantity $f_{HQ}$ determines the amplitude of RCs and indicates the width of the absolute position distribution, where by definition $f_{HQ} = 1$ means diffraction from perfectly flat layers. $P_{HQ}$ determines the shape of RCs and provides the average atomic positions ($z_Q$). The values of $f_{HQ}$ are generally low in Bi2212, which is related to the atomic displacements or supermodulation structure in Bi2212. The very low value of $f_{HQ}$ for O in the CuO layer might be related to more vibrational disorder and the presence of supermodulation.

Supplemental Table S1 The values of $f_{HQ}$, $P_{HQ}$, and the absolute positions ($z_Q$) of the first contributing planes in the unit cell, as determined by fitting the core-level RCs to Eq. (S2). The $c$ lattice constant of Bi2212 is 30.7 Å [S7]. Estimated errors are: $\Delta f_{HQ} = \pm 0.03$, $\Delta P_{HQ} = \pm 0.02$, $\Delta z_Q = d_{002} \times \Delta P_{HQ} = \pm 0.31$ Å.

| Atom | $f_{HQ}$ | $P_{HQ}$ | $z_Q$ (Å) | Atom | $f_{HQ}$ | $P_{HQ}$ | $z_Q$ (Å) |
|---|---|---|---|---|---|---|---|
| Bi | 0.55 | 0.92 | 29.47 | Cu | 0.58 | 0.53 | 23.49 |
| O(Bi) | 0.45 | 0.92 | 29.47 | O(Cu) | 0.21 | 0.60 | 24.56 |
| Ca(LBE) | 0.54 | 0.50 | 23.03 | Sr | 0.36 | 0.67 | 25.63 |
| Ca(HBE) | 0.41 | 0.67 | 25.63 | O(Sr) | 0.43 | 0.77 | 27.17 |

**S4. Valence-band rocking curves: SW modeling based on dynamical diffraction**

In analyzing the valence-band RC of Fig. 4(a) in the main text, we have made the assumption that the matrix elements are primarily controlled by the region near the core, as discussed previously in connection with XPS or HAXPES spectra [S4,S5,S12,S13,S14]. The intensity from a given valence subshell $Qn\ell$ in layer $i$ at depth $z_i$ can then, by analogy with Eq. (S2) be described by

$$I_{VB,Qn\ell,i}(E_b,\theta_{inc}) = \rho_{Qn\ell,i}(E_b) \frac{d\sigma_{Qn\ell}}{d\Omega} \left[ 1 + R(\theta_{inc}) + \sum_i^N \frac{e^{\frac{-z_i}{\Lambda_e \sin\theta_e}}}{I_A} \times 2\cos(2\theta_B)\sqrt{R(\theta_{inc})} f_{HQ} \cos(\varphi_H(\theta_{inc}) - 2\pi P_{HQ}(z_i)) \right], \quad (S5)$$

where $\rho_{Qnl,k}(E_b)$ is the density of states in layer $i$, projected onto $Qn\ell$ character, but assumed not to change with layer, so the $i$ index can be dropped to $\rho_{Qnl}(E_b)$, $d\sigma_{Qnl}(h\nu,\varepsilon)/d\Omega$ is the energy- and polarization- dependent differential photoelectric cross section for subshell $Qn\ell$. The total valence band intensity is thus

$$I_{VB}(E_b,\theta_{inc}) = \sum_{Qn\ell,i} I_{VB,Qn\ell}(E_b,\theta_{inc}) = \sum_{Qn\ell,i} \rho_{Qn\ell}(E_b) \frac{d\sigma_{Qn\ell}}{d\Omega} I_{Qn\ell,i}(\theta_{inc}) \equiv \sum_{Qn\ell,i} D_{Qn\ell}(E_b) I_{Qn\ell,i}(\theta_{inc}), \quad (S6)$$

where we have defined

$$D_{Qn\ell}(E_b) = \rho_{Qn\ell}(E_b)\frac{d\sigma_{Qn\ell}}{d\Omega}, \tag{S7}$$

which will be the experimentally layer-projected quantity; see Eq. (1) in the main text. Now, assuming that the normalized SW effect on a core-level $Qn'\ell'$ in the same layer is the same as that for the $Qn\ell$ valence level, we have from Eqs. (S6) and (S7):

$$I_{VB}(E_b, \theta_{inc}) = \sum_{Qn\ell} I_{VB, Qn\ell}(E_b, \theta_{inc}) = \sum_{Qn\ell} D_{Qn\ell}(E_b)\overline{I}_{Qn'\ell'}(E_b), \tag{S8}$$

which is equivalent to Eq. (1) in the main article. The $Qn\ell$ choices for us are those for which the cross sections are dominant, as described in the text: *Cu 3d* in $CuO_2$, *Sr 4p* in SrO, and *Bi 5d* in BiO, which makes the rocking curves for *Cu 2p, Sr 3d* and *Bi $4f_{7/2}$* the natural choices for the $\overline{I}_{Qn'\ell'}(\theta_{inc})$ in our analysis.

## S5. Photoelectric cross-section-weighted DOSs

As mentioned in the test and the previous section, a given valence spectrum is a linear sum of the individual DOSs $\rho_{Qn\ell}(E_b)$ weighted by matrix elements, or in our assumed high-energy limit, the differential photoelectric cross section $d\sigma_{Qn\ell}/d\Omega$. The differential photoelectric cross section, using the dipole approximation, is given by equation (S9) [S15]:

$$\frac{d\sigma_{Qn\ell}(hv,\varepsilon)}{d\Omega} = \frac{\sigma_{Qn\ell}}{4\pi}[1+\frac{\beta}{2}(3\cos^2\alpha - 1)], \tag{S9}$$

where $\sigma_{Qn\ell}$ is the total photoionization cross section of subshell $Qn\ell$, $\beta$ is the dipole asymmetry parameter, $\alpha$ is the angle between the direction of photoelectron emission and the polarization direction. The experimentally layer-projected DOS $D_i(E_b)$ from the three characteristic atom layers ($i$ = BiO, SrO, and $CuO_2$) approximately equals the cross-section-weighted DOSs summed over the constituting atoms in the same layer. For example, the layer-projected $CuO_2$ DOS is

$$D_{CuO_2}(E_b) \approx \sum_{Cu-n\ell} \frac{d\sigma_{Cu-n\ell}}{d\Omega}\rho_{Cu-n\ell}(E_b) + \sum_{O-n\ell} \frac{d\sigma_{O-n\ell}}{d\Omega}\rho_{O-n\ell}(E_b). \tag{S10}$$

Supplemental Table S2 lists the cross sections and asymmetry parameters of the dominant atomic orbitals in VB intensity for the elements in Bi2212, which are used for calculating the cross-section-weighted DOSs of BiO, SrO, and CuO₂ layers. The values were obtained from Ref. [S16]. The fact that the angle between incidence and polarization is very near the magic-angle for which $3\cos^2\alpha-1$ is zero means that the asymmetry parameter has little influence. Figure S4(a) shows the layer-projected DOSs from each layer calculated by DFT calculations incorporating the known supermodulation structures. Their resulting cross-section-weighted DOSs are shown in Fig. S4(b), which are also presented in Figs. 4(c)(d) of the article.

Supplemental Table S2 The cross sections ($\sigma_{nl}$) and asymmetry parameters ($\beta$) of the dominant atomic orbitals in VB intensity for all the elements in Bi2212.

| Element | Orbital | $\sigma_{nl}$ | $\beta$ | Element | Orbital | $\sigma_{nl}$ | $\beta$ |
|---|---|---|---|---|---|---|---|
| Ca | 4s | 0.0010 | 2.00 | O | 2s | 0.0067 | 2.00 |
| Ca | 3p | 0.0251 | 1.45 | O | 2p | 0.0014 | 0.70 |
| Cu | 4s | 0.0007 | 2.00 | Bi | 6s | 0.0030 | 2.00 |
| Cu | 3p | 0.1003 | 1.60 | Bi | 6p | 0.0028 | 1.67 |
| Cu | 3d | 0.0422 | 1.11 | Bi | 5d | 0.0919 | 1.21 |
| Sr | 5s | 0.0008 | 2.00 | | | | |
| Sr | 4p | 0.0271 | 1.67 | | | | |

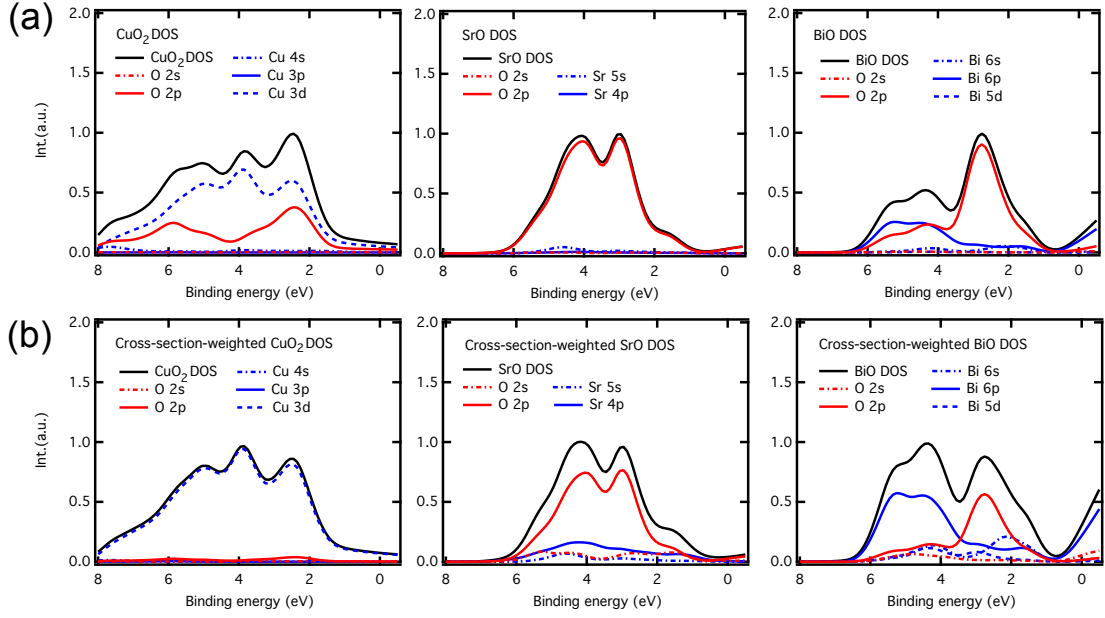

FIG. S4. Photoelectron-cross-section-weighted DOSs. (a) The layer-projected DOSs calculated by DFT calculations incorporating the known supermodulation structures. (b) The resulting layer-projected, cross-section-weighted DOSs calculated using equations (S9) and (S10).

## S6. Atomic coordinates and density functional theory calculations

The atomic coordinates and lattice constants of Bi2212 with and without supermodulation (SM) structure used for our DFT calculations and for our SW dynamical diffraction calculations are listed in Supplemental Table S3 and Table S4, respectively. The unit cell of Bi2212 with supermodulation (SM) structure is $\sqrt{2}a \times \sqrt{2}a \times c$, but the unit cell of Bi2212 without SM is $a \times a \times c$. Therefore, the larger in-plane lattice constants for the Bi2212 with SM leads to a smaller 1st Brillouin zone. The bands of Bi2212 without SM structure are folded in order to directly compare their band structures at the high symmetry points, as shown in Fig. S5.

Supplemental Table S3 The structural properties, such as the atomic coordinates and lattice constants, of ideal Bi2212 for DFT calculations. The space group is I4/mmm and the lattice constants a = b = 3.825 Å, c = 30.82 Å. These values are obtained from Ref. [S17]. x, y, and z are expressed as fractions of a, b, and c. For these coordinates, the center of the unit cell is in the Ca layer.

| Element | Multiplicity | x | y | z |
|---|---|---|---|---|
| Bi | 4 | 0 | 0 | 0.3 |
| Sr | 4 | 0 | 0 | 0.1 |
| Cu | 4 | 0.5 | 0.5 | 0.05 |
| Ca | 2 | 0 | 0 | 0 |
| O(1) | 8 | 0 | 0.5 | 0.05 |
| O(2) | 4 | 0 | 0 | 0.2 |
| O(3) | 4 | 0 | 0 | 0.385 |

Supplemental Table S4 The structural properties, such as the atomic coordinates and lattice constants, of supermodulated Bi2212 for DFT calculations. The space group is Amaa (No. 66), and the lattice constant a = 5.4054 Å, b = 5.4016 Å, c = 30.7152 Å. These values are obtained from Ref. [S7]. x, y, and z are expressed as fractions of a, b, and c. For these coordinates, the center of the unit cell is in between the Bi atoms.

| Element | Multiplicity | x | y | z |
|---|---|---|---|---|
| Bi | 8 | 0.052 | 0.2745 | 0.0524 |
| Sr | 8 | 0 | 0.75 | 0.3597 |
| Cu | 8 | 0.5 | 0.75 | 0.3033 |
| Ca | 4 | 0.5 | 0.25 | 0.25 |
| O(1) | 8 | 0.75 | 0 | 0.201 |
| O(2) | 8 | 0.25 | 0.5 | 0.201 |
| O(3) | 8 | 0 | 0.25 | 0.385 |
| O(4) | 8 | 0.5 | 0.27 | 0.0524 |

We note here that it has also been observed that the SM structure can in fact have a larger period, including up to a fivefold larger unit cell in plane [S10,S18]. We have here only considered the SM in a twofold enlargement of the unit cell, due to the much greater computational effort required to include SM up to the fivefold larger in-plane unit cell, and we believe this should include most of the essential physics.

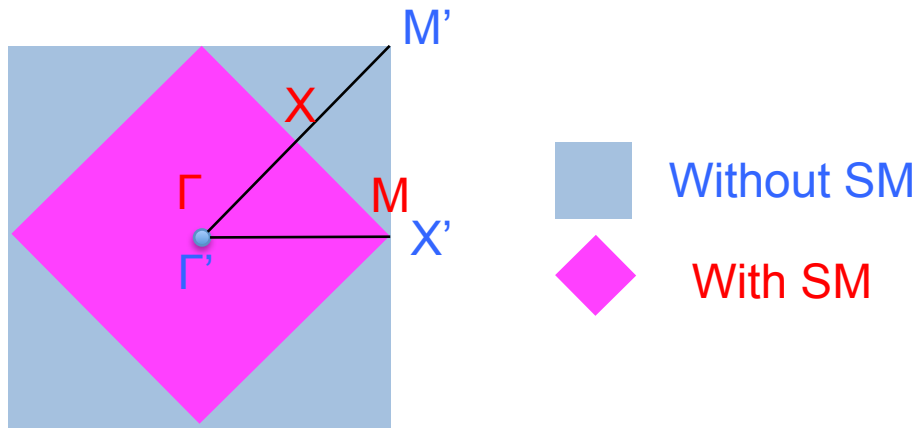

FIG. S5. 1st Brillouin zone of the Bi2212 with and without supermodulation (SM) structure. M', X', and Γ' are the high symmetry points for Bi2212 without SM, while M, X, and Γ are for Bi2212 with SM.